\documentstyle[aps, pra]{revtex}

\begin{document}
\title{Faraday patterns in Bose--Einstein condensates. Amplitude equation for rolls
in the parametrically driven, damped Gross--Pitaevskii equation}
\author{Germ\'{a}n J. de Valc\'{a}rcel\thanks{%
Email address: german.valcarcel@uv.es}}
\address{Departament d'\`{O}ptica, Universitat de Val\`{e}ncia, Dr. Moliner 50,\\
46100-Burjassot, Spain.}
\maketitle

\begin{abstract}
The parametrically driven, damped Gross--Pitaevskii equation, which models
Bose--Einstein condensates in which the interatomic $s-$wave scattering
length is modulated in time, is shown to support spatially modulated states
in the form of rolls. A Landau equation with broken phase symmetry is shown
to govern the dynamics of the roll amplitude.
\end{abstract}

\section{Model}

We consider spontaneous pattern formation in dilute Bose--Einstein
condensates (BEC) whose interatomic $s-$wave scattering length is
periodically varied in time \cite{faraday}. This modulation can be achieved
by different means like, e.g., the use of magnetic \cite{magnetic}, electric 
\cite{electric}, or light \cite{light} fields. When damping is considered 
\cite{damping} such BEC can be described by the following driven, damped
Gross--Pitaevskii (GP) equation, 
\begin{equation}
\partial _{t}\psi \left( {\bf r},t\right) =\left( 1-i\gamma \right) \left[
-\nabla ^{2}\psi +V\left( {\bf r}\right) \psi +\left| \psi \right| ^{2}\psi %
\right] +%
{\textstyle{a\left( t\right) -a_{0} \over a_{0}}}%
\left| \psi \right| ^{2}\psi .
\end{equation}
which has been written in appropriate normalized variables. Parameter $%
\gamma $ accounts for damping (values on the order $0.01-0.1$ seem to be
appropriate for actual BECs \cite{damping}), $a\left( t\right) $ represents
the instantaneous value of the interatomic $s-$wave scattering length and $%
a_{0}$ represents its mean value. We shall assume the simplest case $a\left(
t\right) =a_{0}\left[ 1+2\alpha \cos \left( 2\omega t\right) \right] $.

In the following we shall consider ''pancake'' (2D) or ''cigar'' (1D) shaped
BECs in which the trapping potential $V\left( {\bf r}\right) $ strongly
confines the condensate in one direction or two directions, respectively,
whilst along the other direction(s) it extends sufficiently as compared with
the typical wavelength of the emerging pattern. If a parabolic trapping
potential $V\left( {\bf r}\right) =-1+\frac{1}{2}\left( \omega
_{x}^{2}x^{2}+\omega _{y}^{2}y^{2}+\omega _{z}^{2}z^{2}\right) $ is used
(the arbitrary offset is set to $-1$ for mathematical convenience) the above
situation can be fulfilled whenever $\omega _{z}\gg \omega \gg \omega
_{x},\omega _{y}$ for the 2D case and $\omega _{y},\omega _{z}\gg \omega \gg
\omega _{x}$ for the 1D case. In a first approximation these inequalities
allow: (i) to neglect the confined direction(s) in the descripition of the
BEC dynamics, and (ii) to approximate the potential by a constant. Both
approximations come from the fact that the characteristic wavelength of the
selected pattern turns out to be, at the same time, much smaller than the
size of the weakly confined direction and much larger than the strongly
confined direction. Hence we consider along the rest of this work the
following GP equation \cite{faraday,1D}: 
\begin{equation}
\partial _{t}\psi \left( x,t\right) =\left( 1-i\gamma \right) \left[
-\partial _{x}^{2}\psi -\psi +\left| \psi \right| ^{2}\psi \right] +2\alpha
\cos \left( 2\omega t\right) \left| \psi \right| ^{2}\psi .  \label{GP}
\end{equation}

Eq. (\ref{GP}) admits the following homogeneous state 
\begin{equation}
\psi =\exp \left[ -i\left( \alpha /\omega \right) \sin \left( 2\omega
t\right) \right] ,  \label{ground}
\end{equation}
which, in the absence of modulation ($\alpha =0$) reduces to the BEC ground
state $\psi =1$ (the chemical potential is null in this representation
because of the choice of the offset in the trapping potential).

\section{Linear stability analysis: the parametric resonance}

We wish to know whether the {\it spatially homogeneous} external driving is
able to induce a spontaneous spatial--symmetry breaking of (\ref{ground}).
Fot that we perform next a linear stability analysis of (\ref{ground}) by
adding a small perturbation to that solution in the form $\psi =\exp \left[
-i\left( \alpha /\omega \right) \sin \left( 2\omega t\right) \right] \left[
1+w\left( t\right) \cos \left( kx\right) \right] $. Substitution of this
expression into Eq. (\ref{GP}) and linearization with respect to $w$ leads
to the following coupled equations for $u=%
\mathop{\rm Re}%
w$, and $v=%
\mathop{\rm Im}%
w$: 
\begin{eqnarray}
u^{\prime }\left( t\right)  &=&-\gamma \left( 2+k^{2}\right) u+k^{2}v, \\
v^{\prime }\left( t\right)  &=&-\gamma k^{2}v-\left[ 2+k^{2}+4\alpha \cos
\left( 2\omega t\right) \right] u,
\end{eqnarray}
which can be combined to yield \cite{faraday} 
\begin{equation}
u^{\prime \prime }\left( t\right) +2\gamma \left( 1+k^{2}\right) u^{\prime
}\left( t\right) +\left[ \left( 1+\gamma ^{2}\right) \Omega ^{2}\left(
k\right) +4k^{2}\alpha \cos \left( 2\omega t\right) \right] u\left( t\right)
=0,  \label{mathieu}
\end{equation}
where 
\begin{equation}
\Omega \left( k\right) =k\sqrt{2+k^{2}},
\end{equation}
is the nonlinear dispersion relation for the perturbations.

Eq. (\ref{mathieu}) is a Mathieu equation with damping, analogous to that
describing parametrically driven, damped pendula, and ubiquitous in the
description of parametric forcing \cite{cross}. Its solutions are well known
which, according to Floquet's Theorem, can be written as 
\begin{equation}
u\left( t\right) =%
\mathop{\rm Re}%
f\left( t\right) e^{\mu t},  \label{floquet}
\end{equation}
where $f$ is a periodic complex function of period $\pi /\omega $ and $\mu $
is the so called Floquet exponent (in the case of the Mathieu equation $\mu
/i$ is known as Mathieu characteristic exponent). The BEC ground state (\ref
{ground}) will be unstable against perturbations with wavenumber $k$
whenever $%
\mathop{\rm Re}%
\mu >0$. A general property of Eq. (\ref{mathieu}) is that $%
\mathop{\rm Re}%
\mu >0$ within a series of resonance ''tongues'' very much like the
parametric resonances observed in liquids vibrated vertically \cite{cross}.
For small $\gamma $ and $\alpha $ (the cases we consider here) these tongues
are located, as a function of $\Omega $, around $\Omega \left( k_{n}\right)
=n\omega $, $n=1,2,3,\ldots $ Within these tongues the BEC ground state
undergoes a spontaneous spatial-symmetry breaking and perturbations with
wavenumber $k_{n}=\sqrt{\sqrt{1+n^{2}\omega ^{2}}-1}$ amplify. The Floquet
exponents can be numerically determined by using standard mathematical
methods \cite{mathematica}. Useful analytical information can be obtained in
the limit of weak damping $\gamma \ll 1$. Taking into account that for $%
\gamma =\alpha =0$ the solution to (\ref{mathieu}) is of the form $u\left(
t\right) =%
\mathop{\rm Re}%
Ue^{i\Omega t}$, if we consider $\alpha \sim \left| \omega -\Omega \right|
\sim \gamma $, a perturbative expression for $u$ can be obtained by allowing 
$U$ to be a slowly varying function of time with the result 
\begin{eqnarray}
u\left( t\right)  &=&%
\mathop{\rm Re}%
\left[ U_{+}\exp \left( \lambda _{+}t\right) +U_{-}\exp \left( \lambda
_{+}t\right) \right] e^{i\Omega t},  \label{uaprox1} \\
\lambda _{\pm } &=&i\left( \omega -\Omega \right) -\gamma \left(
1+k^{2}\right) \pm \sqrt{\left( \frac{\alpha k^{2}}{\Omega }\right)
^{2}-\left( \omega -\Omega \right) ^{2}}.  \label{uaprox2}
\end{eqnarray}
Note that (\ref{uaprox1}) can be written as (\ref{floquet}) with $f\left(
t\right) =\exp \left( i2\omega t\right) $, and 
\begin{equation}
\mu =-i\omega -\gamma \left( 1+k^{2}\right) +\sqrt{\left( \frac{\alpha k^{2}%
}{\Omega }\right) ^{2}-\left( \omega -\Omega \right) ^{2}}.
\end{equation}
Finally the condition $%
\mathop{\rm Re}%
\mu \geq 0$ reads 
\begin{equation}
\alpha \geq \frac{\Omega }{k^{2}}\sqrt{\gamma ^{2}\left( 1+k^{2}\right)
^{2}+\left( \omega -\Omega \right) ^{2}},  \label{tongue1}
\end{equation}
where the equality defines the boundary (neutral stability line) of the
first resonance tongue. Note that (\ref{tongue1}) indicates that, for fixed $%
\omega $, the threshold for pattern formation is minimum at $\Omega =\omega $%
, i.e., for wavenumbers $k=k_{1}=\sqrt{\sqrt{1+\omega ^{2}}-1}$. The minimum
parametric driving is hence predicted to be $\alpha _{\min }^{n=1}\left(
\omega \right) =\gamma \omega \sqrt{1+\omega ^{2}}/\left( \sqrt{1+\omega ^{2}%
}-1\right) $. Following a similar analysis for the second resonance tongue ($%
\Omega \simeq 2\omega $) in the limit $\alpha \sim \left| \omega -\Omega
\right| \sim \gamma ^{1/2}$, it follows that $u\left( t\right) $ can be
written as (\ref{uaprox1}) with 
\begin{equation}
\lambda _{\pm }=i\left( 2\omega -\Omega \right) -\gamma \left(
1+k^{2}\right) \pm \frac{1}{3}\sqrt{20\left( \frac{\alpha ^{2}k^{4}}{\Omega
^{3}}\right) ^{2}-24\frac{\alpha ^{2}k^{4}}{\Omega ^{3}}\left( 2\omega
-\Omega \right) -9\left( 2\omega -\Omega \right) ^{2}},
\end{equation}
from which the second tongue ($%
\mathop{\rm Re}%
\lambda _{+}\geq 0$) runs: 
\begin{equation}
\alpha \geq \frac{\Omega ^{3/2}}{k^{2}}\sqrt{\frac{3}{10}}\sqrt{\sqrt{%
5\gamma ^{2}\left( 1+k^{2}\right) ^{2}+9\left( 2\omega -\Omega \right) ^{2}}%
+2\left( 2\omega -\Omega \right) }.
\end{equation}
The minimum driving amplitude needed in this case, $\alpha _{\min }^{n=2}$,
occurs at $\Omega =2\omega $. In this case $\alpha _{\min }^{n=2}\sim \sqrt{%
\gamma }$ hence the second resonance tongue is excited at larger driving
amplitudes than the first one, for which $\alpha _{\min }^{n=1}\sim \gamma $%
. Note in both cases that for $\gamma =0$ (no damping) the threshold for
both (and in fact any) resonance tongues is $\alpha =0$ hence all are
simultaneously excited at vanishingly small values of driving amplitude.
Damping hence breaks this degeneracy and selects the first resonance tongue
at low drivings.

\section{The roll pattern: amplitude equation}

In the following we study the simplest pattern supported by the GP equation (%
\ref{GP}) under the previously described parametric instability. We consider
a roll pattern in the form
\begin{equation}
\psi \left( x,t\right) =e^{-i\left( \alpha /\omega \right) \sin \left(
2\omega t\right) }\left[ 1+\varepsilon w_{1}\left( t\right) \cos \left(
kx\right) +\varepsilon ^{2}w_{2}\left( x,t\right) +\varepsilon
^{3}w_{3}\left( x,t\right) +O\left( \varepsilon ^{4}\right) \right] ,
\label{roll}
\end{equation}
where $0<\varepsilon \ll 1$ is an auxiliary small parameter. In order to
deal with a small roll component (of order $\varepsilon $) we assume that
the amplitude of the parametric driving is also small, say $\alpha =O\left(
\varepsilon ^{2}\right) $ (that this is the proper scaling for $\alpha $ is
justified a posteriori by the consistency of the final result). On the other
hand for the parametric excitation to be effective Eq. (\ref{tongue1}) must
be fulfilled. If we want to take into account both the resonance condition
and the effect of damping we must impose $\gamma =O\left( \varepsilon
^{2}\right) $ and $\left( \omega -\Omega \right) =O\left( \varepsilon
^{2}\right) $. Summarizing we consider in the following the scalings 
\begin{equation}
\omega =\Omega +\varepsilon ^{2}\omega _{2},\quad \alpha =\varepsilon
^{2}\alpha _{2},\quad \gamma =\varepsilon ^{2}\gamma _{2}.  \label{scalings}
\end{equation}
Our goal is to find an equation for the roll complex amplitude $w_{1}$. This
will be done using a standard multiple timescale technique \cite
{cross,newell}. For this purpose we introduce a slow time
\begin{equation}
\tau =\varepsilon ^{2}t,  \label{tau}
\end{equation}
and allow all coefficients of the expansion to depend formally both on $t$
and $\tau $: 
\begin{eqnarray}
w_{1}\left( t\right)  &=&u_{11}\left( t,\tau \right) +iv_{11}\left( t,\tau
\right) ,  \label{slow1} \\
w_{j}\left( x,t\right)  &=&u_{j}\left( x,t,\tau \right) +iv_{j}\left(
x,t,\tau \right) .  \label{slow2}
\end{eqnarray}
Finally Eqs. (\ref{roll}), (\ref{scalings}), (\ref{slow1}) and (\ref{slow2})
are introduced into the GP equation (\ref{GP}). After using the chain rule
for differentiation $\partial _{t}\rightarrow \partial _{t}+\varepsilon
^{2}\partial _{\tau }$ and equating equal powers in $\varepsilon $ an
infinite hierarchy of differential equations is obtained.

\subsection{Order $\protect\varepsilon ^{1}$}

This is the first nontrivial order and reads

\begin{eqnarray}
\partial _{t}u_{11}\left( t,\tau \right) -k^{2}v_{11}\left( t,\tau \right) 
&=&0, \\
\partial _{t}v_{11}\left( t,\tau \right) +\left( 2+k^{2}\right) u_{11}\left(
t,\tau \right)  &=&0,
\end{eqnarray}
whose solution can be written as 
\begin{eqnarray}
u_{11}\left( t,\tau \right)  &=&\left[ r\left( \tau \right) e^{i\Omega t}+%
\overline{r}\left( \tau \right) e^{-i\Omega t}\right] ,  \label{u11} \\
v_{11}\left( t,\tau \right)  &=&i\frac{\Omega }{k^{2}}\left[ r\left( \tau
\right) e^{i\Omega t}-\overline{r}\left( \tau \right) e^{-i\Omega t}\right] ,
\label{v11}
\end{eqnarray}
where $r\left( \tau \right) $ stands for a yet arbitrary complex function of
the slow time, and the overbar denotes complex conjugation.

\subsection{Order $\protect\varepsilon ^{2}$}

At this order we find

\begin{eqnarray}
\partial _{t}u_{2}\left( x,t,\tau \right) +\partial _{x}^{2}v_{2}\left(
x,t,\tau \right)  &=&\left[ 1+\cos \left( 2kx\right) \right] u_{11}\left(
t,\tau \right) v_{11}\left( t,\tau \right) , \\
\partial _{t}v_{2}\left( x,t,\tau \right) +\left( 2-\partial _{x}^{2}\right)
u_{2}\left( x,t,\tau \right)  &=&-%
{\textstyle{1 \over 2}}%
\left[ 1+\cos \left( 2kx\right) \right] \left[ 3u_{11}^{2}\left( t,\tau
\right) +v_{11}^{2}\left( t,\tau \right) \right] ,
\end{eqnarray}
whose solution can be written as 
\begin{eqnarray}
u_{2}\left( x,t,\tau \right)  &=&u_{20}\left( t,\tau \right) +u_{22}\left(
t,\tau \right) \cos \left( 2kx\right) , \\
v_{2}\left( x,t,\tau \right)  &=&v_{20}\left( t,\tau \right) +v_{22}\left(
t,\tau \right) \cos \left( 2kx\right) ,
\end{eqnarray}
where each of the coefficients verifies 
\begin{eqnarray}
\partial _{t}u_{20}\left( t,\tau \right)  &=&i\frac{\Omega }{k^{2}}\left[
r^{2}\left( \tau \right) e^{i2\Omega t}-\overline{r}^{2}\left( \tau \right)
e^{-i2\Omega t}\right] , \\
\partial _{t}v_{20}\left( t,\tau \right)  &=&-2u_{20}\left( t,\tau \right)
-2\left( 2+k^{-2}\right) \left| r\left( \tau \right) \right| ^{2}+\left(
k^{-2}-1\right) \left[ r^{2}\left( \tau \right) e^{i2\Omega t}\overline{r}%
^{2}\left( \tau \right) e^{-i2\Omega t}\right] , \\
\partial _{t}u_{22}\left( t,\tau \right)  &=&4k^{2}v_{22}\left( t,\tau
\right) +i\frac{\Omega }{k^{2}}\left[ r^{2}\left( \tau \right) e^{i2\Omega
t}-\overline{r}^{2}\left( \tau \right) e^{-i2\Omega t}\right] , \\
\partial _{t}v_{22}\left( t,\tau \right)  &=&-\left( 2+4k^{2}\right)
u_{22}\left( t,\tau \right) -2\left( 2+k^{-2}\right) \left| r\left( \tau
\right) \right| ^{2}+\left( k^{-2}-1\right) \left[ r^{2}\left( \tau \right)
e^{i2\Omega t}\overline{r}^{2}\left( \tau \right) e^{-i2\Omega t}\right] ,
\end{eqnarray}
which solved yield 
\begin{eqnarray}
u_{20}\left( t,\tau \right)  &=&-2\left( 1+k^{-2}\right) \left| r\left( \tau
\right) \right| ^{2}+\frac{1}{2k^{2}}\left[ r\left( \tau \right) e^{i\Omega
t}+\overline{r}\left( \tau \right) e^{-i\Omega t}\right] ^{2},  \label{u20}
\\
u_{22}\left( t,\tau \right)  &=&-\frac{1}{2k^{2}}\left[ r\left( \tau \right)
e^{i\Omega t}+\overline{r}\left( \tau \right) e^{-i\Omega t}\right] ^{2},
\label{u22} \\
v_{20}\left( t,\tau \right)  &=&v_{200}\left( \tau \right) +\frac{i}{2\Omega 
}\left[ r^{2}\left( \tau \right) e^{i2\Omega t}-\overline{r}^{2}\left( \tau
\right) e^{-i2\Omega t}\right] ,  \label{v20} \\
v_{22}\left( t,\tau \right)  &=&-\frac{i\Omega }{2k^{4}}\left[ r^{2}\left(
\tau \right) e^{i2\Omega t}-\overline{r}^{2}\left( \tau \right) e^{-i2\Omega
t}\right] ,  \label{v22}
\end{eqnarray}
where $v_{200}\left( \tau \right) $ is a yet undetermined function of the
slow time, which is not fixed by the present analysis. This information
should be obtainable by extending the calculation up to higher orders of the
expansion. Anyway, as will be seen below, the knowledge of $v_{200}\left(
\tau \right) $ is not relevant for our purposes.

\subsection{Order $\protect\varepsilon ^{3}$. The amplitude equation}

This is the last order we will consider. It reads

\begin{eqnarray}
\partial _{t}u_{3}\left( x,t,\tau \right) +\partial _{x}^{2}v_{3}\left(
x,t,\tau \right)  &=&f_{u}\left( t,\tau \right) \cos \left( kx\right)
+g_{u}\left( t,\tau \right) \cos \left( 3kx\right) ,  \label{or3u} \\
\partial _{t}v_{3}\left( x,t,\tau \right) +\left( 2-\partial _{x}^{2}\right)
u_{3}\left( x,t,\tau \right)  &=&f_{v}\left( t,\tau \right) \cos \left(
kx\right) +g_{v}\left( t,\tau \right) \cos \left( 3kx\right) ,  \label{or3v}
\end{eqnarray}
where
\begin{eqnarray}
f_{u} &=&-\partial _{\tau }u_{11}-\gamma _{2}\left( 2+k^{2}\right)
u_{11}+\left( 2v_{20}+v_{22}\right) u_{11}+\left( 2u_{20}+u_{22}\right)
v_{11}+%
{\textstyle{3 \over 4}}%
\left( u_{11}^{2}+v_{11}^{2}\right) v_{11},  \label{fu} \\
f_{v} &=&-\partial _{\tau }v_{11}-\gamma _{2}k^{2}v_{11}-4\alpha _{2}\cos
\left( 2\Omega t+2\omega _{2}\tau \right) u_{11}-3\left(
2u_{20}+u_{22}\right) u_{11}-\left( 2v_{20}+v_{22}\right) v_{11}-%
{\textstyle{3 \over 4}}%
\left( u_{11}^{2}+v_{11}^{2}\right) u_{11}  \label{fv} \\
g_{u} &=&u_{11}v_{22}+u_{22}v_{11}+%
{\textstyle{1 \over 4}}%
\left( u_{11}^{2}+v_{11}^{2}\right) v_{11}, \\
g_{v} &=&-3u_{11}u_{22}-v_{11}v_{22}-%
{\textstyle{1 \over 4}}%
\left( u_{11}^{2}+v_{11}^{2}\right) u_{11}.
\end{eqnarray}
The solution to Eqs. (\ref{or3u}) and (\ref{or3v}) can be written as 
\begin{eqnarray}
u_{3}\left( x,t,\tau \right)  &=&u_{31}\left( t,\tau \right) \cos \left(
kx\right) +u_{33}\left( t,\tau \right) \cos \left( 3kx\right) , \\
v_{3}\left( x,t,\tau \right)  &=&v_{31}\left( t,\tau \right) \cos \left(
kx\right) +v_{33}\left( t,\tau \right) \cos \left( 3kx\right) ,
\end{eqnarray}
where each of the coefficients verifies 
\begin{eqnarray}
\partial _{t}u_{31}\left( t,\tau \right) -k^{2}v_{31}\left( t,\tau \right) 
&=&f_{u}\left( t,\tau \right) ,  \label{s1} \\
\partial _{t}v_{31}\left( t,\tau \right) +\left( 2+k^{2}\right) u_{31}\left(
t,\tau \right)  &=&f_{v}\left( t,\tau \right) ,  \label{s2} \\
\partial _{t}u_{33}\left( t,\tau \right) -9k^{2}v_{33}\left( t,\tau \right) 
&=&g_{u}\left( t,\tau \right) , \\
\partial _{t}v_{33}\left( t,\tau \right) +\left( 2+9k^{2}\right)
u_{33}\left( t,\tau \right)  &=&g_{v}\left( t,\tau \right) .
\end{eqnarray}
The last two equations do not give us relevant information for our purposes.
On the contrary Eqs. (\ref{s1}) and (\ref{s2}) determine the evolution
equation for the complex amplitude of the roll $r$. This comes from the fact
that these equations contain resonant terms which yield divergent solutions
unless a solvability condition is imposed. This is clearly seen by writing
Eqs. (\ref{s1}) and (\ref{s2}) in vector form: 
\begin{equation}
\partial _{t}\left[ 
\begin{array}{c}
u_{31}\left( t,\tau \right)  \\ 
v_{31}\left( t,\tau \right) 
\end{array}
\right] =\left[ 
\begin{array}{cc}
0 & k^{2} \\ 
-\left( 2+k^{2}\right)  & 0
\end{array}
\right] \left[ 
\begin{array}{c}
u_{31}\left( t,\tau \right)  \\ 
v_{31}\left( t,\tau \right) 
\end{array}
\right] +\left[ 
\begin{array}{c}
f_{u}\left( t,\tau \right)  \\ 
f_{v}\left( t,\tau \right) 
\end{array}
\right] .
\end{equation}
Upon diagonalizing this equation we obtain 
\begin{equation}
\partial _{t}\chi \left( t,\tau \right) =-i\Omega \chi \left( t,\tau \right)
+f\left( t,\tau \right) ,  \label{chi}
\end{equation}
(and its complex conjugate) where 
\begin{eqnarray}
\chi \left( t,\tau \right)  &=&\left( 2+k^{2}\right) u_{31}\left( t,\tau
\right) +i\Omega v_{31}\left( t,\tau \right) , \\
f\left( t,\tau \right)  &=&\left( 2+k^{2}\right) f_{u}\left( t,\tau \right)
+i\Omega f_{v}\left( t,\tau \right) .  \label{f}
\end{eqnarray}
Eq. (\ref{chi}) can only be solved if the driving term $f$ does not contain
elements oscillating as $\exp \left( -i\Omega t\right) $. Upon substituting
Eqs. (\ref{fu}) and (\ref{fv}) into Eq. (\ref{f}), and making use of Eqs. (%
\ref{u11}), (\ref{v11}) and (\ref{u20})--(\ref{v22}), the solvability
condition is found to be
\begin{equation}
\frac{dr}{d\tau }=-\gamma _{2}\left( 1+k^{2}\right) r+i\frac{\alpha _{2}k^{2}%
}{\Omega }e^{2i\omega _{2}\tau }\overline{r}-i\frac{3+5k^{2}}{\Omega }\left|
r\right| ^{2}r.  \label{drdtau}
\end{equation}
Finally we define a new roll complex amplitude
\begin{equation}
R\left( t\right) =\varepsilon r\left( \tau \right) e^{-i\omega _{2}\tau },
\label{R}
\end{equation}
and turn back to the original parameters by undoing the scalings (\ref
{scalings}) with the result
\begin{equation}
\frac{dR}{dt}=-\left[ \gamma \left( 1+k^{2}\right) +i\left( \omega -\Omega
\right) \right] R+i\frac{\alpha k^{2}}{\Omega }\overline{R}-i\frac{3+5k^{2}}{%
\Omega }\left| R\right| ^{2}R.  \label{landau}
\end{equation}
Eq. (\ref{landau}) is the searched roll amplitude equation. It is a Landau
equation with broken phase symmetry (note the presence of the linear term
proportional to $\overline{R}$). The roll solution (\ref{roll}) can be
written in terms of $R$ making use of Eqs. (\ref{slow1}), (\ref{u11}), (\ref
{v11}) and (\ref{R}). To the leading order the roll reads
\begin{eqnarray}
\psi \left( x,t\right)  &=&e^{-i\left( \alpha /\omega \right) \sin \left(
2\omega t\right) }\left[ 1+w\left( t\right) \cos \left( kx\right) \right] ,
\label{rolldef} \\
w\left( t\right)  &=&\left( 1-\Omega /k^{2}\right) R\left( t\right)
e^{i\omega t}+\left( 1+\Omega /k^{2}\right) \overline{R}\left( t\right)
e^{-i\omega t}.
\end{eqnarray}

Let us finally note that a straightforward linear stability analysis of the
trivial solution $R=0$ (hence $w=0$) of Eq. (\ref{landau}) yields the same
neutral stability curve given in (\ref{tongue1}).

\begin{center}
Acknowledgment
\end{center}

I thank very useful discussions with Kestutis Staliunas (PTB, Braunschweig,
Germany) and Stefano Longhi (Polit\'{e}cnico di Milano, Milan, Italy). This
work has been supported by the Spanish DGES through project PB98-0935-C03-02.


\begin{references}
\bibitem{faraday}  K. Staliunas, S. Longhi, and G. J. de Valc\'{a}rcel, {\it %
Faraday patterns in Bose--Einstein condensates} (submitted, 2002).

\bibitem{magnetic}  S. Inouye, M. R. Andrews, J. Stenger, H.-J. Miesner,
D.M. Stamper-Kuhn, and W. Ketterle, Nature {\bf 392}, 151 (1998).

\bibitem{electric}  M. Marinescu and L. You, Phys. Rev. Lett. {\bf 81}, 4596
(1988).

\bibitem{light}  P. O. Fedichev, Yu. Kagan, G. V. Shlyapnikov, and J. T. M.
Walraven, Phys. Rev. Lett. {\bf 77}, 2913 (1996).

\bibitem{damping}  We use a very simple, yet compatible with experiments,
phenomenological description of damping. See S. Choi, S. A. Morgan, and K.
Burnett, Phys. Rev. A {\bf 57}, 4057 (1998).

\bibitem{1D}  Eq. (\ref{GP}) holds for 1D condensates, as well as for 2D
condensates whenever 1D patterns are considered, which is the concern of
this work.

\bibitem{cross}  M. C. Cross and P. C. Hohenberg, Rev. Mod. Phys. {\bf 65},
851 (1993).

\bibitem{mathematica}  In particular the Floquet exponents of the Mathieu
equation (\ref{mathieu}) can be computed from the {\tt Mathematica} built-in
function {\tt MathieuCharacteristicExponent}.

\bibitem{newell}  A. C. Newell and J. A. Whitehead, J. Fluid. Mech. {\bf 38}%
, 279 (1969).
\end{references}
\end{document}